\documentclass[letterpaper, conference]{ieeeconf}
\IEEEoverridecommandlockouts
\usepackage[english]{babel}
\usepackage[utf8x]{inputenc}

\usepackage{latexsym,amsmath,amssymb,amsfonts,graphicx,graphics,epsfig,epsf,lscape,subfigure}
\usepackage{float}
\usepackage[section]{placeins}
\usepackage{bm}

\usepackage{wrapfig}
\usepackage{amssymb}

\usepackage{breqn}
\usepackage{dsfont}
\usepackage{pgf}

\usepackage{enumerate}

\def \etal {\emph{et al.}}

\newtheorem{theorem}{Theorem}
\newtheorem{lemma}{Lemma}
\newtheorem{proposition}{Proposition}

\newtheorem{example}{Example}
\newtheorem{remark}{Remark}

\usepackage[font=footnotesize]{caption}

\def \bs {\boldsymbol}
\def \mc {\mathcal}

\def\expt{\mathbb{E}}
\def\real{\mathbb{R}}

\def\naturals{\mathbb{N}}
\newcommand{\until}[1]{\{1,\dots, #1\}}

\newcommand{\setdef}[2]{\{#1 \; | \; #2\}}
\newcommand{\seqdef}[2]{\{#1\}_{#2}}

\newcommand{\slfrac}[2]{\left.#1\middle/#2\right.}
\newcommand{\lnn}[1]{%
	\ln  \left(#1\right)%
}
\newcommand{\expp}[1]{%
	\exp \! \left(#1\right)%
}

\newcommand\oprocendsymbol{\hbox{$\square$}}
\newcommand\oprocend{\relax\ifmmode\else\unskip\hfill\fi\oprocendsymbol}

\renewcommand{\theenumi}{(\roman{enumi}}

\newcommand\bit[1]{\textit{\textbf{#1}}}

\renewcommand{\baselinestretch}{0.985}

\title{Distributed Cooperative Decision-Making in Multiarmed Bandits: Frequentist and Bayesian Algorithms
\thanks{This revision provides a correction to the original paper, which appeared in the Proceedings of the 2016 IEEE Conference on Decision and Control (CDC).  The second statement of Proposition 1 and Theorem 1 are new from \cite{arxiv:LandgrenSL15_new} and Lemma 1 is new.  These are used to prove regret bounds in Theorems 2 and 3.}
\thanks{This research has been supported by ONR grant  N00014-14-1-0635, ARO grant W911NF-14-1-0431, and the DoD through the NDSEG Program.}}

\author{Peter Landgren, Vaibhav Srivastava, and Naomi Ehrich Leonard
\thanks{P. Landgren and N. E. Leonard are with the Department of Mechanical and Aerospace Engineering, Princeton University, Princeton, NJ, USA,\tt{ \{landgren, naomi\}@princeton.edu}.}
\thanks{V. Srivastava is with the Department of Electrical and Computer Engineering, Michigan State University, East Lansing, MI, USA, \tt{vaibhav@egr.msu.edu}}
}

\begin{document}
\maketitle
\setlength{\abovedisplayskip}{5.85pt}
\setlength{\belowdisplayskip}{5.85pt}

\begin{abstract}
We study distributed cooperative decision-making under the explore-exploit tradeoff in the multiarmed bandit (MAB) problem. We extend state-of-the-art frequentist and Bayesian algorithms for single-agent MAB problems to cooperative distributed algorithms for multi-agent MAB problems in which agents communicate according to a fixed network graph.  We rely on a running consensus algorithm for each agent's estimation of mean rewards from its own rewards and the estimated rewards of its neighbors.  We prove the performance of these algorithms and show that they asymptotically recover the performance of a centralized agent.   Further, we rigorously characterize the influence of the communication graph structure on the decision-making performance of the group. 
\end{abstract}

\section{Introduction}

Cooperative decision-making under uncertainty is ubiquitous in natural systems as well as in engineering networks. 
%
A fundamental feature of decision-making under uncertainty is the \emph{explore-exploit} tradeoff:
the decision-making agent needs to learn the unknown system parameters (exploration), while maximizing its parameter-dependent decision-making objective (exploitation). 

Multiarmed bandit (MAB) problems are canonical formulations of the explore-exploit tradeoff. In a stochastic MAB problem a set of options (arms) is given. A stochastic reward with an unknown mean is associated with each option. A player can pick only one option at a time, and the player's objective is to maximize the cumulative expected reward over a sequence of choices. In an MAB problem, the player needs to balance the tradeoff between learning the mean reward at each arm (exploration), and picking the arm with maximum mean reward (exploitation). 

MAB problems are pervasive across a variety of scientific communities and have found application in diverse areas including control and robotics~\cite{VS-PR-NEL:14, MYC-JL-FSH:13},  ecology~\cite{JRK-AK-PT:78, VS-PR-NEL:13}, and communications~\cite{anandkumar2011distributed}. 
Despite the prevalence of the MAB problem, the research on MAB problems has primarily focused on policies for a single agent. The increasing importance of networked systems warrants the development of distributed algorithms for multiple communicating agents faced with MAB problems. 
In this paper, we build upon previous work by extending two popular single-agent algorithms for the stochastic MAB problem to the distributed multiple agent setting and analyze decision-making performance as a function of the network structure.

The MAB problem has been extensively studied (see~\cite{SB-NCB:12} for a survey). In their seminal work, Lai and Robbins~\cite{lai1985asymptotically} established a logarithmic lower bound on  the expected number of times a sub-optimal arm needs to be selected by an optimal policy in a frequentist setting. In another seminal work, Auer~\etal~\cite{PA-NCB-PF:02} developed the upper confidence bound (UCB) algorithm for the stochastic MAB problem, which achieves the lower bound in~\cite{lai1985asymptotically} uniformly in time.  

The MAB problem has also been studied in the Bayesian setting. 
Kaufmann~\etal~\cite{EK-OC-AG:12} proposed the Bayes-UCB algorithm and showed that it achieves Lai-Robbins' lower bound for Bernoulli rewards and uninformative priors. 
Reverdy~\etal~\cite{reverdy2014modeling} developed and analyzed the upper credible limit (UCL) algorithm for correlated multiarmed bandits by applying the approach of~\cite{EK-OC-AG:12}  to the case of Gaussian rewards. 


The classical single-agent MAB problem was extended by Anantharam~\etal~\cite{VA-PV-JW:87} to the setting {of a single-agent with multiple plays. }
Recently,  researchers~\cite{kalathil2014decentralized,gai2014distributed,anandkumar2011distributed} have studied the decentralized multi-player MAB problem with no communication among agents. 
Kar~\etal~\cite{kar2011bandit} investigated the multi-agent MAB problem in a leader-follower setting. 
%
Here, we use a running consensus algorithm~\cite{PB-SM-VM:08} for assimilation of information.
Running consensus, {also known as dynamic consensus,} has been used to study  related collective
decision-making models in social networks~\cite{VS-NEL:13f}. 


In the present paper we study the distributed cooperative MAB problem in which agents are faced with a stochastic MAB problem and communicate their information with their neighbors in an undirected and connected communication graph. We use a set of running consensus algorithms for cooperative estimation of the mean reward at each arm, and we design an arm selection heuristic that leads to an order-optimal performance for the group. The major contributions of this paper are as follows.

First, we propose and thoroughly analyze the coop-UCB2 and coop-UCL algorithms. We derive bounds on decision-making performance for the group and characterize the influence of the network structure on performance. 
To predict  nodal performance, we propose a measure of nodal  ``explore-exploit centrality,'' which depends on the location of the node in the graph.

Second, we demonstrate that the ordering of nodes by performance predicted by our explore-exploit centrality measure matches  the order obtained using numerical simulations. We also show that the incorporation of priors that are well-informative about the correlation structure 
markedly improve performance.

The remainder of the paper is organized as follows.  In Section~\ref{sec:problem} we introduce the cooperative MAB problem. 
In Section~\ref{sec:coop-est} we recall a cooperative estimation algorithm. We review the coop-UCB algorithm in Section~\ref{DistributedDecisionMaking}, and propose and analyze the improved coop-UCB2 and new coop-UCL algorithms. We illustrate our analytic results with numerical examples in Section~\ref{NetworkPerformanceAnalysis}. We conclude in Section~\ref{FinalRemarks}.

\section{Cooperative Multiarmed Bandit Problem}\label{sec:problem}
Consider an MAB problem with $N$ arms and $M$ decision-making agents. 
The reward associated with arm $i\in \until{N}$ is a random variable with an unknown mean $m_i$. 
Let the communication topology of agents be modeled 
by an undirected graph $\mathcal{G}$ in which each node represents an agent and edges represent the communication between agents. 
Let $A\in \real^{M\times M}$ be the adjacency matrix associated with $\mc G$ and let $L \in \real^{M\times M}$ be the corresponding Laplacian matrix. We assume that the graph $\mc G$ is connected, i.e., there exists a path between each pair of nodes. 

Let the $k$-th agent choose arm $i^k(t)$ at time $t \in \until{T}$ and receive a reward $r^k(t)$. 
The objective of each decision-maker $k$ is to choose using its local information a sequence of arms $\seqdef{i^k(t)}{t\in \until{T}}$  such that the total expected cumulative reward $\sum_{k=1}^M\sum_{t=1}^T m_{i^k(t)}$ is maximized, where $T$ is the horizon length of the sequential allocation process.

For an MAB problem, the expected \emph{regret} of agent $k$ at time $t$ is defined by $R^k(t) = m_{i^*}-m_{i^k(t)}$, where $m_{i^*} = \max \setdef{m_i}{i\in\until{N}}$. The collective objective of the $M$ decision-makers can be equivalently defined as minimizing the expected cumulative regret defined by $\sum_{k=1}^M \sum_{t=1}^T R^k(t) = \sum_{k=1}^M \sum_{i=1}^N \Delta_i \expt[n_{i}^k(T)]$, 
where $n_{i}^k(T)$ is the cumulative number of times arm $i$ has been chosen by agent $k$ until time $T$ and $\Delta_i = m_{i^*}-m_i$ is the expected regret due to picking arm $i$ instead of arm $i^*$. It is known that the regret of any algorithm for an MAB problem is asymptotically lower bounded by a logarithmic function of the horizon length $T$~\cite{lai1985asymptotically}, \cite{VA-PV-JW:87}, i.e., no algorithm can achieve an expected cumulative regret smaller than a logarithmic function of horizon length as $T \to \infty$. 

In this paper, we focus on Gaussian rewards, i.e., the reward at arm $i$ is sampled from a Gaussian distribution with mean $m_i$ and variance $\sigma_s^2$.  We assume that the variance $\sigma_s^2$ is known and is the same at each arm. In the context of Gaussian rewards, the lower bound~\cite{VA-PV-JW:87} on the expected number of times  a suboptimal arm $i$ is selected by a fusion center that has access  to reward for each agent is
\begin{equation}
\sum_{k=1}^M \expt[n_i^k(T)] \ge \left( \frac{2 \sigma_s^2}{\Delta_i^2} +o(1) \right) \ln T. \label{eqn:fusioncenterregret}
\end{equation}

\noindent
In the following, we will design policies that sample a suboptimal arm $i$ within a constant factor of the above bound.

\section{Cooperative Estimation of Mean Rewards} \label{sec:coop-est}
In this section we recall the algorithm for cooperative estimation of mean rewards proposed in our earlier work~\cite{arxiv:LandgrenSL15_new,arxiv:LandgrenSL15}.
%

\subsection{Cooperative Estimation Algorithm}
For distributed cooperative estimation of the mean reward at each arm $i$, we employ two running consensus algorithms to estimate (i) total reward provided at the arm, and (ii)  the total number of times the arm has been sampled. 

Let $\hat{s}_i^k(t)$ and $\hat{n}_i^k(t)$ be agent $k$'s estimate of the total reward provided at arm $i$ per unit agent and the total number of times arm $i$ has been selected until time $t$ per unit agent, respectively. Using $\hat{s}_i^k(t) $ and  $\hat{n}_i^k(t) $ agent $k$ can calculate $\hat{\mu}_i^k(t)$, the estimated empirical mean of arm $i$ at time $t$ as 
\begin{equation}
\hat{ \mu}_i^{k}(t) = \frac{\hat{s}_i^{k}(t)}{\hat{n}_i^{k}(t)}.  \label{eqnmean}
\end{equation}

Let $i^k(t)$ be the arm sampled by agent $k$ at time $t$ and let $\xi_i^k(t) = \mathds{1} (i^k(t) =i)$.   $\mathds{1}(\cdot)$ is the indicator function, here equal to 1 if $i^k(t)= i$ and 0 otherwise. 
{For simplicity of notation we define $r_i^k(t)$ as the realized reward at arm $i$ for agent $k$, which is a random variable sampled from $\mathcal{N}(m_i,\sigma_s^2)$, and the corresponding accumulated reward is $r^k(t) = r_i^k(t) \cdot \mathds{1} (i^k(t) =i)$.}
Let $P$ be a row stochastic matrix given by
\begin{equation} 
P= \mathcal{I}_M - \frac{\kappa}{d_{\text{max}}} L, \label{Pdefn}
\end{equation}
where $\mathcal{I}_M$ is the identity matrix of order $M$, $\kappa \in (0,1]$ is a step size parameter \cite{olfati2004consensus}, $d_{\text{max}} =\max \setdef{\text{deg}(k)}{k \in \until{M}}$, and $\text{deg}(k)$ is the degree of node $k$.  

The estimates $\hat{n}_i^k(t)$ and $\hat{s}_i^k(t)$are updated locally using running consensus~\cite{PB-SM-VM:08} as follows:
\begin{align} \label{nhatdefn}
\mathbf{\hat{n}}_i(t) &= P \mathbf{\hat{n}}_i(t-1) + P\boldsymbol{\xi}_i(t), \\
\text{and} \quad \mathbf{\hat{s}}_i(t) &= P \mathbf{\hat{s}}_i(t-1) + P(\mathbf{r}_i(t) \circ \boldsymbol{\xi}_i(t)), \label{shatdefn}
\end{align}
where $\mathbf{\hat{ n}}_i(t)$, $\mathbf{\hat{ s}}_i(t)$, $\boldsymbol{\xi}_i(t)$, and $\mathbf{r}_i(t)$ are vectors of $\hat n_i^k(t)$, $\hat s_i^k(t)$, $\xi_i^k(t)$, and and $r_i^k(t)$, $k\in \until{M}$, respectively, { and $\circ$ denotes element-wise multiplication (Hadamard product).

\subsection{Analysis of the Cooperative Estimation Algorithm}
We now recall the performance of the estimation algorithm defined by~(\ref{eqnmean}--\ref{shatdefn}). Let $n_i^{\text{cent}}(t) \equiv \frac{1}{M} \sum_{\tau=1}^t \mathbf{1}_M^\top \boldsymbol{\xi}_i(\tau)$ be the total number of times arm $i$ has been selected per unit agent until time $t$, and let $s_i^{\text{cent}}(t) \equiv \frac{1}{M} \sum_{\tau=1}^t \boldsymbol{\xi}_i^\top(\tau) \mathbf{r}_i(\tau)$ be the
total reward provided at arm $i$ per unit agent until time $t$.  Also, let $\lambda_i$ denote the $i$-th largest eigenvalue of $P$, $\mathbf{u}_i$  the eigenvector corresponding to $\lambda_i$, $u_i^d$  the $d$-th entry of $\mathbf{u}_i$, and
\begin{equation}
\epsilon_n =  \sqrt{M}  \sum_{p=2}^M \frac{|\lambda_p|}{1-|\lambda_p|}.  \label{epsilonndef}
\end{equation}

{Note that $\lambda_1=1$ and $\mathbf{u}_1 = \mathbf{1}_M/\sqrt{M}$. Let us define
\begin{align*}
\nu_{pj}^{\text{+sum}} &= \sum_{d=1}^M  u_p^d u_j^d \mathds{1}(u_p^k u_j^k \geq 0) \\
\text{and} \quad \nu_{pj}^{\text{-sum}} &=  \sum_{d=1}^M u_p^d u_j^d \mathds{1}(u_p^k u_j^k \leq 0). 
\end{align*}
We also define
\begin{equation}
a_{pj}(k) =  \begin{cases}
\nu_{pj}^{\text{+sum}}u_p^k u_j^k, & 
\! \! \! \! \text{if } \lambda_p \lambda_j \geq 0 \; \&\; u_p^k u_j^k  \geq 0, \\
\nu_{pj}^{\text{-sum}}u_p^k u_j^k, &
\! \! \! \! \text{if }  \lambda_p \lambda_j \geq 0 \; \&\; u_p^k u_j^k  \leq 0, \\
\nu_{pj}^{\text{max}} |u_p^k u_j^k | , & \! \! \! \! \text{if }  \lambda_p \lambda_j < 0,
\end{cases}\label{apjdefn}
\end{equation}
where $\nu_{pj}^{\text{max}} = \max{\{ |\nu_{pj}^{\text{-sum}}|,  \nu_{pj}^{\text{+sum}}\}} $.}  Furthermore, let 
\begin{equation}
\epsilon_c^k   =    M \sum_{p=1}^M \sum_{j=2}^M \frac{|\lambda_p \lambda_j| }{1-|\lambda_p \lambda_j|} a_{pj}(k). \label{eqn:epsilonck_defn}
\end{equation}

We note that both $\epsilon_n$ and $\epsilon_c^k$ depend only on the topology of the communication graph and are measures of distributed cooperative estimation performance.
We now recall the following results from~\cite{arxiv:LandgrenSL15_new,arxiv:LandgrenSL15}.

\begin{proposition}[\bit{Performance of cooperative estimation}]\label{prop:coop-est}
For the distributed estimation algorithm defined in~(\ref{eqnmean}--\ref{shatdefn}), 
the following statements hold
\begin{enumerate}
	\item the estimate $\hat n_i^k(t)$ satisfies
	\begin{align*}
	n_i^{\text{cent}}(t) - \epsilon_n \le  \hat{n}_i^k(t) \le   n_i^{\text{cent}}(t) + \epsilon_n;
	\end{align*}
	\item {the following inequality holds for the estimate   $\hat {n}_i^k(t)$ and the sequence $\seqdef{\xi_i^j(\tau)}{\tau\in \until{t}}$, $ j \in \until{M}$
	\begin{equation*}
	\sum_{\tau=1}^{t} \sum_{j=1}^M \left(\sum_{p=1}^M \lambda_p^{t-\tau+1}  u_p^k u_p^j  \right)^2 \xi_i^j(\tau) \leq \frac{ \hat n_i^k(t) + \epsilon_c^k}{M}.
	\end{equation*}}
\end{enumerate}
\end{proposition}

{
\begin{theorem}[\bit{Estimator Deviation Bounds}] \label{Thm:EstDevBoudnsCondensed}
	For the estimates 	$\hat{s}_i^k(t)$ and $\hat{n}_i^k(t)$ obtained using equations~\eqref{nhatdefn} and \eqref{shatdefn}, the following concentration inequality holds 
	\begin{equation}
	\mathbb{P}\Bigg( \! \frac{\hat{s}_i^k(t) \!-\! m_i \hat{n}_i^k(t)}{\left(\frac{1}{M} \left(\hat{n}_i^k(t) \!+\! \epsilon_c^k\right)\right)^{\slfrac{1}{2}}} \! > \! \delta \! \Bigg) \!<\! \Bigg\lceil \!\frac{\lnn{t \!+\! \epsilon_n}}{\lnn{1\!+\!\eta}}\! \Bigg\rceil  \!\expp{\frac{-\delta^2}{2\sigma_s^2} G(\eta) \! \! } ,
	\end{equation}
	where 	$\delta > 0$, $\eta >0$, $G(\eta) = (1-\slfrac{\eta^2}{16})$, and $\epsilon_n$ and $\epsilon_c^k$  are defined in ~\eqref{epsilonndef} and ~\eqref{eqn:epsilonck_defn}, respectively. 
\end{theorem}
\begin{proof}
	See ~\cite{arxiv:LandgrenSL15_new,arxiv:LandgrenSL15}.
\end{proof}}

\section{Frequentist Cooperative Decision-Making}
\label{DistributedDecisionMaking}
In this section, we first review the coop-UCB algorithm proposed in our earlier work~\cite{arxiv:LandgrenSL15_new,arxiv:LandgrenSL15}.  We then improve on this algorithm with a new algorithm: coop-UCB2. Unlike coop-UCB the improved algorithm  does not require each agent to know the global graph structure.
Finally, we compute bounds on the performance of the group for this algorithm as a function of the graph structure.

\subsection{The coop-UCB Algorithm}
The  coop-UCB algorithm is analogous to the UCB algorithm~\cite{PA-NCB-PF:02}, and uses a modified decision-making heuristic that captures the effect of the additional information an agent receives through communication with other agents as well as the rate of information propagation through the network. 

The coop-UCB algorithm is initialized by each agent sampling each arm once and proceeds as follows.  { At time $t$ each agent $k$ selects the arm with maximum $Q_i^{k}(t-1) = \hat{\mu}_i^{k}(t-1) + C_i^k(t-1)$, where
%
\begin{equation}
C_i^k(t-1) = \sigma_s \;\sqrt[]{\frac{2 \gamma}{G(\eta)} \cdot \frac{\hat{n}_i^{k}(t-1) +  \epsilon_c^k}{M\hat{n}_i^{k}(t-1)}\cdot\frac{ \lnn{ t-1}}{\hat{n}_i^{k}(t-1)}}, \label{Cdefn}
\end{equation}
$\gamma>1$, $\eta \in (0,4)$, and $G(\eta) = 1-\slfrac{\eta^2}{16}$.} Then, at each time $t$, each agent $k$ updates its cooperative estimate of the mean reward at each arm using the distributed cooperative estimation algorithm described in~(\ref{eqnmean}--\ref{shatdefn}).

The coop-UCB provides a distributed, cooperative solution to the MAB problem such that every agent in the network achieves logarithmic regret.   However, the heuristic $Q_i^k$ may be overly restrictive in the sense that it requires the agent $k$ to know $\epsilon_c^k$, which depends on the global graph structure. Further, agents with a relatively high $\epsilon_c^k$ are essentially forced to explore more while better positioned agents exploit, leading to wide disparities in performance across some networks.  We will develop the coop-UCB2 algorithm that addresses these issues in the next section.

\subsection{The coop-UCB2 Algorithm}

{ The coop-UCB2 algorithm is initialized by each agent sampling each arm once and proceeds as follows. At time $t$ each agent $k$ selects the arm with maximum $Q_i^{k}(t-1) = \hat{\mu}_i^{k}(t-1) + C_i^k(t-1)$, where
\begin{equation}
C_i^k(t-1) = \sigma_s \;\sqrt[]{\frac{2\gamma}{G(\eta)} \cdot \frac{\hat{n}_i^{k}(t-1) +  f(t-1)}{M\hat{n}_i^{k}(t-1)}\cdot\frac{ \lnn{t-1}}{\hat{n}_i^{k}(t-1)}}, \label{C2defn}
\end{equation}
 $f(t)$ is an increasing sublogarthmic function, $\gamma>1$, $\eta \in (0,4)$, and $G(\eta) = 1-\slfrac{\eta^2}{16}$. Then, at each time $t$, each agent $k$ updates its cooperative estimate of the mean reward at each arm using the distributed cooperative estimation algorithm described in~(\ref{eqnmean}--\ref{shatdefn}).} Note that the heuristic $Q_i^k$ requires the agent $k$ to know the total number of agents $M$, but not the global graph structure. 

\begin{theorem}[\bit{Regret of the coop-UCB2 Algorithm}]\label{thm:regret-coop-ucb2}
For the coop-UCB2 algorithm and the cooperative Gaussian MAB problem, 
the number of times a suboptimal arm $i$ is selected by all agents until time $T$ satisfies
{
	\begin{multline*}
	\sum_{k=1}^M\mathbb{E}[n_i^{k}(T)]  \! \leq 2\sum_{k=1}^M  (t_k^\dagger\!-\!1)  + \max \bigg\{M, \\
	\bigg\lceil M \epsilon_n + \frac{4 \sigma_s^2 \gamma \ln T}{\Delta_i^2 G(\eta)} \Big( 1 + \sqrt{1 + \frac{\Delta_i^2 M G(\eta)}{2\gamma \sigma_s^2} \frac{f(T)}{\ln T}} \Big)
 \bigg\rceil \bigg\} \\
+ \!\frac{2M}{\lnn{1\!+\!\eta}}  \! \bigg( \frac{1}{(\gamma - 1)^2}  \!+\! {\frac{\gamma \lnn{1\!+\!\epsilon_n)(1\!+\!\eta)}}{\gamma - 1} +1}\bigg),
	\end{multline*}
where $t^\dagger_k= f^{-1}(\epsilon_c^k)$. 	
}

\end{theorem}
%
\noindent {\bf Proof:}
We proceed similarly to~\cite{PA-NCB-PF:02}.  The number of selections of a suboptimal arm $i$ by all agents until time $T$ is 
\begin{align}
&\sum_{k=1}^M  n_i^k(T) 
\leq  \sum_{k=1}^M {\sum_{t=N+1}^{T}}  \mathds{1}(Q_i^k(t-1) \geq Q_{i^*}^k(t-1)) {+ M} \nonumber \\
& \leq \! {\max\{M,  A\}  +\! \sum_{k=1}^M \sum_{t=N}^{T-1} \mathds{1}(Q_i^k(t) \geq Q_{i^*}^k(t), M {n}^{\text{cent}}_i(t) \geq A),}  \label{suboptimal-samples}
\end{align}
where $A >0$ is a constant that will be chosen later.  

At a given time $t+1$ an individual agent $k$ will choose a suboptimal arm only if 
$ Q_i^k(t) \geq Q_{i^*}^k(t)$. 
For this condition to be true at least one of the following three conditions must hold:
\begin{align}
\hat{\mu}_{i^*}^k(t) &\leq m_{i^*} - C_{i^*}^k(t) \label{1stcond} \\
\hat{\mu}_{i}^k(t) &\geq m_{i} + C_{i}^k(t) \label{2ndcond} \\
m_{i^*} &< m_{i} + 2 C_{i}^k(t). \label{3rdcond} 
\end{align}

We now bound the probability that \eqref{2ndcond} holds using Theorem \ref{Thm:EstDevBoudnsCondensed}: 
\begin{align*}
&\mathbb{P}\left( {\eqref{2ndcond}} \textrm{ holds } | \, t \geq  t_k^\dagger \right) \\
& \qquad = \mathbb{P} \!\left(\! \frac{\hat{s}_{i}^k(t) - m_{i} \hat{n}_{i}^k(t)}{\sqrt{ \frac{1}{M} \left( \hat{n}_{i}^{k}(t) +  f(t)\right) }} \! \geq \! \sigma_s \sqrt{\frac{2 \gamma \lnn{t}}{G(\eta)}} \, \Bigg | \, t \geq t_k^\dagger \right)\\
& \qquad \leq \mathbb{P} \!\left(\! \frac{\hat{s}_{i}^k(t) - m_{i} \hat{n}_{i}^k(t)}{\sqrt{ \frac{1}{M} \left( \hat{n}_{i}^{k}(t) +  \epsilon_c^k\right) }} \! \geq \! \sigma_s \sqrt{\frac{2 \gamma \lnn{t}}{G(\eta)}} \, \Bigg | \, t \geq t_k^\dagger \right)\\
& \qquad \leq \left( \frac{\lnn{t}}{\lnn{1+\eta}} + \frac{\lnn{1+\epsilon_n}}{\lnn{1+\eta}} + 1\right) \frac{1}{t^\gamma}.
\end{align*}

It also follows analogously that 
\begin{equation*}
 \mathbb{P}\left( {\eqref{1stcond}} \textrm{ holds } | t \geq  t_k^\dagger \right) \leq \left( \frac{\lnn{t}}{\lnn{1+\eta}} + \frac{\lnn{1+\epsilon_n}}{\lnn{1+\eta}} + 1\right) \frac{1}{t^\gamma}.
\end{equation*}

We now examine the event \eqref{3rdcond}. 
\begin{align}
m_{i^*} & < m_i + 2C_i^k(t) \nonumber \\ 
\implies \hat{n}_i^k(t)^2 \frac{\Delta_i^2M G(\eta)}{8 \sigma_s^2} &- \gamma \hat{n}_i^k(t)\ln(t) -  \gamma f(t) \ln(t) < 0. \label{eqn:3rdcondquad}
\end{align}
The quadratic equation~\eqref{eqn:3rdcondquad} can be solved to find the roots, and if $\hat{n}_i^k(t)$ is greater than the larger root the inequality will never hold.
Solving the quadratic equation~\eqref{eqn:3rdcondquad}, we obtain that event~\eqref{3rdcond} does not hold if
\begin{align*}
\hat{n}_i^k(t) &\geq \frac{4 \sigma_s^2 \gamma \ln(t)}{\Delta_i^2MG(\eta)} \!+\! \sqrt{\Big(\frac{4 \gamma \sigma_s^2 \ln(t)}{\Delta_i^2MG(\eta)}\Big)^2 \!+  \frac{8 \sigma_s^2f(t) \gamma \ln(t)}{\Delta_i^2 MG(\eta)}  } \\
& = \frac{4 \sigma_s^2 \gamma \ln t}{\Delta_i^2 MG(\eta)} \bigg( 1 + \sqrt{1 + \frac{\Delta_i^2 MG(\eta)}{2\sigma_s^2 \gamma } \frac{f(t)}{\ln t}} \bigg).
\end{align*}

Now, we set $A = \Big\lceil M \epsilon_n + \frac{4 \sigma_s^2 \gamma \ln T}{\Delta_i^2 G(\eta)} \big( 1 + \sqrt{1 + \frac{\Delta_i^2 M G(\eta)}{2\gamma \sigma_s^2} \frac{f(T)}{\ln T}} \big)
 \Big\rceil$. It follows from monotonicity of $f(t)$ and $\ln(t)$ and statement (i) of Proposition~\ref{prop:coop-est}
  that event~\eqref{3rdcond} does not hold if $M n^{\text{cent}}_i(t) > A$.  
  
  Therefore, from \eqref{suboptimal-samples} we see that
  \begin{align*}\label{eq:total-suboptimal}
	&\sum_{k=1}^M  \mathbb{E} \left[ n_i^k(T)\right] \leq {\max\{M, A\} }+ 2\sum_{k=1}^M  \sum_{t=1}^{t_k^\dagger-1} 1\\
	& \quad \qquad + \!\frac{2}{\lnn{1\!+\!\eta}} \sum_{k=1}^M \sum_{t=t_k^\dagger}^T \! \left( \frac{ \lnn{t}}{t^\gamma} \!+\! \frac{\lnn{(1\!+\!\epsilon_n)(1\!+\!\eta)}}{t^\gamma} \right) \\
	&\quad \leq {\max\{M, A\} }+ 2\sum_{k=1}^M  (t_k^\dagger-1)\\
	&\quad \qquad + \!\frac{2M}{\lnn{1\!+\!\eta}} \sum_{t=1}^T \! \left( \frac{ \lnn{t}}{t^\gamma} \!+\! \frac{\lnn{(1\!+\!\epsilon_n)(1\!+\!\eta)}}{t^\gamma} \right) \\
	&\quad \leq {\max\{M, A\} } \!+\! 2\sum_{k=1}^M  (t_k^\dagger\!-\!1) + \!\frac{2M}{\lnn{1\!+\!\eta}}  \! \Big( \frac{1}{(\gamma - 1)^2} \\
	& \qquad \qquad \qquad \!+\! {\frac{\gamma \lnn{1\!+\!\epsilon_n)(1\!+\!\eta)}}{\gamma - 1} +1}\Big)\!,
  \end{align*}
completing the theorem.
\oprocend

\begin{remark}[\bit{Asymptotic Regret for coop-UCB2}]
 In the limit $t\rightarrow + \infty$, $\frac{f(t)}{\ln(t)} \to 0^+$, $\eta \rightarrow 0$, and   
 \vspace{-0.05in}
	\begin{equation*}
		\sum_{k=1}^M \mathbb{E}[n_i^k(T)] \leq  \Big( \frac{8 \sigma_s^2 \gamma}{\Delta_i^2} + o(1) \Big)\ln T  , 
	\end{equation*}
	 \vspace{-0.1in}

\noindent
and we recover the upper bound on regret for a centralized agent as given in \eqref{eqn:fusioncenterregret} within a constant factor. 
\oprocend
\end{remark}

\begin{remark}[\bit{Performance of Individual Agents}] \label{Remark:Indiv}
Theorem \ref{thm:regret-coop-ucb2} provides bounds on the performance of the group as a function of the graph structure, and the logarithmic portion of the bound is independent of agent location.  However, the constant factor is dependent on the agent's position in the network since it depends on $\epsilon_c^k$.  In this sense, $\epsilon_c^k$ can be thought of as a measure of ``explore-exploit centrality,'' which indicates that agents with a higher $\epsilon_c^k$ will contribute more to the group's regret. \oprocend
\end{remark}

\section{Bayesian Cooperative Decision-Making}
\label{sec:PriorsandCorr}
In this section, we extend the coop-UCB2 algorithm to a Bayesian setting and develop the coop-UCL algorithm. The Bayesian setting allows us to model correlated bandits and incorporate a priori knowledge about reward and correlation structure in the Bayesian prior. We first recall the UCL algorithm proposed in~\cite{reverdy2014modeling,EK-OC-AG:12} and  extend it to the cooperative setting. We then analyze the performance of this algorithm for an uninformative prior. 

\subsection{The UCL Algorithm}
The UCL algorithm developed in \cite{reverdy2014modeling} applies the approach of Bayes-UCB~\cite{EK-OC-AG:12} to correlated Gaussian bandits. The UCL algorithm at each time computes the posterior distribution of mean rewards at each option and then computes the $(1-1/{Kt^a})$ upper-credible-limit for each arm, i.e., an upper bound that holds with probability $(1-1/{Kt^a})$ where {$K=\sqrt{2 \pi e}$, $\gamma>1$, and $a = 4/3 \gamma$.} The algorithm chooses the arm with highest upper credible limit. 
For Gaussian rewards, the $(1-1/{Kt^a})$ upper-credible-limit can be written as
\begin{equation} 
Q_i(t) = \nu_i(t) + \sigma_i(t) \Phi^{-1}(1-1/{Kt^a}), \label{eqn:UCL}
\end{equation}
where $\nu_i(t)$ is the posterior mean and  $\sigma_i(t) $  the posterior standard deviation of mean reward at time $t$.  $\Phi^{-1}(\cdot)$ is the standard Gaussian inverse cumulative distribution function.

Let the  prior on rewards from each arm be multivariate Gaussian with mean vector $\bs \nu_0 \in \real^N$ and covariance matrix $\Sigma_0 \in \mathbb{R}^{N \times N}$.  
Then, the posterior mean and covariance of mean reward at time $t$ can be computed using the following recursive update rule~\cite{SMK:93}:
\begin{align}\label{eqn:nu_update}
\begin{split}
\mathbf{q}(t) &= \frac{r(t) \boldsymbol{\phi}(t)}{\sigma_s^2} + \Lambda(t-1) \boldsymbol{\nu}(t-1) \\
\Lambda(t) &= \frac{\boldsymbol{\phi}(t)^\top  \boldsymbol{\phi}(t)}{\sigma_s^2} + \Lambda(t-1), \quad  
\Sigma(t) = \Lambda(t)^{-1}  \\
\boldsymbol{\nu}(t) &= \Sigma(t) \boldsymbol{q}(t),
\end{split}
\end{align} 
where $\boldsymbol{\phi}(t)$  and $\boldsymbol{\nu(t)}$ are column vectors of $\phi_i(t)$ and $\nu_i(t)$, respectively, and $\phi_i(t)$ is the indicator function of selecting arm $i$ at time $t$. The update equation~\eqref{eqn:nu_update} can be reduced to 
\begin{align}\label{eqn:nu_update2}
\begin{split}
\boldsymbol{\nu}(t) &= (\Lambda_0 + \Gamma(t)^{-1})^{-1}(\Gamma(t)^{-1}\boldsymbol{\mu}(t) + \Lambda_0 \boldsymbol{\nu}_0)  \\
\Lambda(t) &= \Lambda_0 + \Gamma(t)^{-1}, \quad \Sigma(t) =(\Lambda(t))^{-1},
\end{split}
\end{align}
where $\Lambda_0 = \Sigma_0^{-1}$, $\Gamma(t)$ is a diagonal matrix with  entries $\frac{\sigma_s^2}{n_i(t)}$, and $\bs \mu(t)$ is the vector of $\mu_i(t)$, which is the empirical mean of rewards from arm $i \in \until{N}$ until time $t$. Note that diagonal entries of $\Sigma(t)$ are $(\sigma_i(t))^2, \; i\in\until{N}$.
 
%

\subsection{The coop-UCL Algorithm}
We now extend the UCL algorithm to the distributed cooperative setting and propose the coop-UCL algorithm. In the coop-UCL algorithm, each agent first computes an approximate posterior distribution of mean rewards conditioned on rewards obtained by all the agents. To this end, each agent uses the approximate frequentist estimator $\hat \mu_i^k$ from Section~\ref{sec:coop-est} in update equation~\eqref{eqn:nu_update2}. 

Let the prior of agent $k$ be a multivariate Gaussian distribution with mean $\bs \nu_0^k$ and covariance $\Sigma_0^k$. Let  $\hat{\Sigma}^k(t)$ and $\boldsymbol{\hat{\nu}}^k(t)$ be the estimated covariance matrix and posterior mean at time $t$, respectively.  Then, the coop-UCL algorithm performs cooperative approximate Bayesian estimation:
\begin{align}\label{eqn:nu_update_coop}
\begin{split}
\boldsymbol{\hat{\nu}}^k(t) &= (\Lambda_0^k + \Gamma^k(t)^{-1})^{-1}(\Gamma^k(t)^{-1}\hat{\boldsymbol{\mu}}^k(t) + \Lambda_0^k \boldsymbol{\nu}_0^k) \\
\hat{\Lambda}^k(t) &= \Lambda_0^k + \Gamma^k(t)^{-1}, \quad {\hat{\Sigma}}^k(t) =(\hat{\Lambda}^k(t))^{-1},
\end{split}
\end{align} 

\noindent
 where $\Gamma^k(t)$ is  a diagonal matrix with diagonal entries ${\sigma_s^2}/{M \hat n_i^k(t)}, \; i \in \until{N}$, and $\Lambda_0^k = (\Sigma_0^k)^{-1}$. 

%
%
%

After computing $\boldsymbol{\hat{\nu}}^k(t-1)$ and $\hat{\Sigma}^k(t-1)$, the coop-UCL algorithm at time $t$ requires each agent $k$ to choose the option with maximum $(1-\alpha(t))$-upper-credible-limit given by
\begin{equation}
Q_i^k(t-1) = \hat{\nu}_i^k(t-1) + \hat{\sigma}_i^k(t-1) \Phi^{-1}(1-\alpha(t-1)), \label{eqn:CoopUCL}
\end{equation}
where $\alpha(t)$ is defined such that
\begin{equation*}
\Phi^{-1}(1-\alpha(t)) = \sqrt{\frac{\hat{n}_i^k(t) + f(t)}{G(\eta) \hat{n}_i^k(t)}} \Phi^{-1}\Big(1- \frac{1}{Kt^a} \Big),
\end{equation*}
 where $\hat{\nu}_i^k(t)$ is the $i$-th entry of $\boldsymbol{\hat{\nu}}^k(t)$, $(\hat{\sigma}_i^k(t))^2$ is the $i$-th diagonal entry of $\hat{\Sigma}^k(t)$, $K = \sqrt{2 \pi e}$, $\gamma > 1$, and $a=4/3 \gamma$. 

\subsection{Regret of the coop-UCL Algorithm}

We now derive bounds on the expected cumulative regret for each agent using the coop-UCL algorithm with uninformative priors for each agent. For an uninformative prior, $\Lambda_0^k =0$, for each $k\in \until{M}$, and consequently, $\bs \hat{\nu}^k(t) = \bs \hat{\mu}^k(t)$ and $\hat{\Sigma}^k(t) = \Gamma^k(t)$.  In addition, we first present a bound on $\Phi^{-1}(\cdot)$.
\begin{lemma}[\bit{Inverse Gaussian CDF Bounds}]\label{lem:ineq}
	For  the standard normal random variable $z$ and the associated inverse cumulative distribution function $\Phi^{-1}(\cdot)$, the following hold for any $\alpha \in [0, 0.5]$, $t\in \naturals$ and $a>1$:
		\begin{equation*}
		\Phi^{-1}(1- \alpha) \le \sqrt{-2 \log (\alpha)} 
		\end{equation*}
		\begin{equation*}
		\Phi^{-1}(1-\alpha)   > \sqrt{-\log(2\pi \alpha^2(1-\log(2\pi \alpha^2)))}
		\end{equation*}
		\begin{equation*}
		\Phi^{-1}\Big(1- \frac{1}{\sqrt{2 \pi e} t^a}\Big) > \sqrt{ \nu \log t^a},
		\end{equation*}
		for $0 < \nu \le 1.59$.
\end{lemma}
\noindent
{\bf Proof:}
The first inequality can be found in~\cite{MA-IAS:64}. The second inequality was established in ~\cite{reverdy2014modeling}. 
%
To establish the last inequality, it suffices to show that
\[
- \log \left( \frac{1}{e t^2} \left( 1 -  \log \left( \frac{1}{e t^2} \right) \right)\right) - \nu \log t  \ge 0,
\]
for $0< \nu \le 1.59$. The left hand side of the above inequality is 
\[
g(t) := 1 - \log 2 + (2- \nu) \log t - \log(1+\log t). 
\]
It can be verified that $g$ admits a unique minimum at $t = e^{(\nu -1)/(2-\nu)}$ and the minimum value is $\nu -\log 2 + \log(2-\nu)$, which is positive for $0< \nu \le 1.59$. 
\oprocend

In the following, we select $\nu =3/2$. 


\begin{theorem}[\bit{Regret of the coop-UCL Algorithm}]\label{thm:regret-coop-ucl}
	For the Gaussian MAB problem and the coop-UCL algorithm with uninformative priors for each agent, 
	the number of times a suboptimal arm $i$ is selected by all agents until time $T$ satisfies
	{\begin{multline*}
	\sum_{k=1}^M\mathbb{E}[n_i^{k}(T)]  \! \leq    2 \!\sum_{k=1}^M  (t_k^\dagger\!-\!1) \! + \! 
	\max \bigg\{ M, \bigg \lceil M\epsilon_n    \\
	 + \bigg(\! \frac{4  \sigma_s^2 \ln{KT^a}}{\Delta_i^2 M G(\eta)} \bigg(\! 1 \!+ \!\sqrt{1 \! + \! \frac{\Delta_i^2 MG(\eta)}{2 \sigma_s^2} \frac{f(T)}{\ln KT^a}} \bigg)\! \bigg) \!  \bigg \rceil \bigg\} \\
+	 \!\frac{2M}{\lnn{1\!+\!\eta}}  \! \left( \frac{1}{(\gamma - 1)^2} \!+\! \frac{\gamma \lnn{1\!+\!\epsilon_n)(1\!+\!\eta)}}{\gamma - 1} +1\right) 
	\end{multline*}}
	
\noindent where 
	$t^\dagger_k= f^{-1}(\epsilon_c^k)$. 

\end{theorem}

%
\noindent
{\bf Proof:}
For uninformative priors, coop-UCL is analogous to coop-UCB2 with $C_i^k(t) = \hat{\sigma}_i^k(t) \Phi^{-1}(1 - \alpha(t))$, {where $\hat{\sigma}_i^k(t) = \sigma_s /\sqrt{M \hat n_i^k(t)}$}. Similar to the proof of Theorem~\ref{thm:regret-coop-ucb2}, we first note that for { {\eqref{2ndcond}} simple manipulations lead to 
{
\begin{align}
\frac{\hat{s}^k_{i}(t) - m_{i}  \hat{n}_{i}^k(t)}{\sqrt{\hat{n}_{i}^k(t) +f(t)}} &\geq \frac{\sigma_s }{\sqrt{M G(\eta)}} \Phi^{-1}\left(1-\frac{1}{Kt^a}\right)\nonumber \\
& > \frac{\sigma_s }{\sqrt{M G(\eta)}} \sqrt{\frac{3a}{2} \ln{t}} \label{eqn:phiupperbound} \\
& = \frac{\sigma_s }{\sqrt{M G(\eta)}} \sqrt{2 \ln{t^\gamma}} \nonumber
\end{align}}
where \eqref{eqn:phiupperbound} follows from Lemma \ref{lem:ineq} for $K=\sqrt{2 \pi e}$.

Using Theorem \ref{Thm:EstDevBoudnsCondensed} we get that 
\begin{align*}
&\mathbb{P}\left( {\eqref{2ndcond}} \textrm{ holds } | \, t \geq  t_k^\dagger \right) \\
& \qquad \leq \mathbb{P} \!\left(\! \frac{\hat{s}_{i}^k(t) - m_{i} \hat{n}_{i}^k(t)}{\sqrt{ \frac{1}{M} \left( \hat{n}_{i}^{k}(t) +  f(t)\right) }} \! \geq \! \sigma_s \sqrt{\frac{2 \gamma \lnn{t}}{G(\eta)}} \, \Bigg | \, t \geq t_k^\dagger \right)\\
& \qquad \leq \left( \frac{\lnn{t}}{\lnn{1+\eta}} + \frac{\lnn{1+\epsilon_n}}{\lnn{1+\eta}} + 1\right) \frac{1}{t^\gamma}
\end{align*}
resulting in sub-logarithmic regret as in Theorem \ref{thm:regret-coop-ucb2}.}

We now examine the event \eqref{3rdcond}.  Following the argument in the proof of Theorem~\ref{thm:regret-coop-ucb2} and using the upper bound on $\Phi^{-1}(\cdot)$ { from Lemma \ref{lem:ineq},} we obtain that the event~\eqref{3rdcond} does not hold if 
{
\begin{align*}
\hat{n}_i^k(t) &\geq 
\frac{4 \sigma_s^2  \ln{Kt^a}}{\Delta_i^2 M G(\eta)} \bigg( 1 + \sqrt{1 + \frac{\Delta_i^2 M G(\eta)}{2\sigma_s^2 } \frac{f(t)}{\ln{Kt^a}}} \bigg).
\end{align*}

We set $A \! = \! \Big\lceil M \epsilon_n \! + \! \frac{4 \sigma_s^2  \ln KT^a}{\Delta_i^2 M G(\eta)} \Big( \! 1 +  \sqrt{1 + \frac{\Delta_i^2 M G(\eta)}{2\sigma_s^2 } \frac{f(T)}{\ln KT^a}} \Big) \! \Big\rceil $} and the theorem follows by proceeding similarly to the proof of  Theorem~\ref{thm:regret-coop-ucb2}. 
\oprocend

\section{Numerical Illustrations}
\label{NetworkPerformanceAnalysis}

\begin{figure}[]

\centering
\subfigure[]{\label{5AgentGraphSim}

		\includegraphics[width=.8\columnwidth]{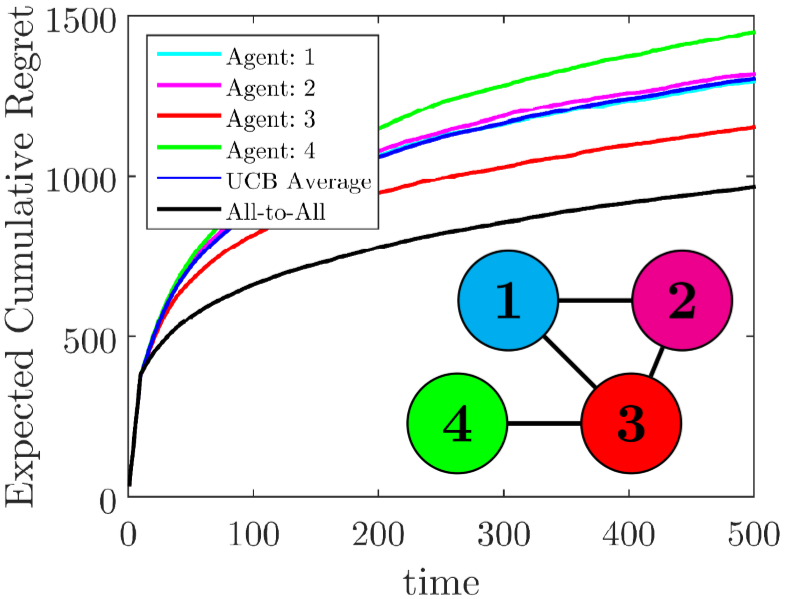}}\\

\subfigure[]{ \label{MethodCompare} 
	\includegraphics[width=.8\columnwidth]{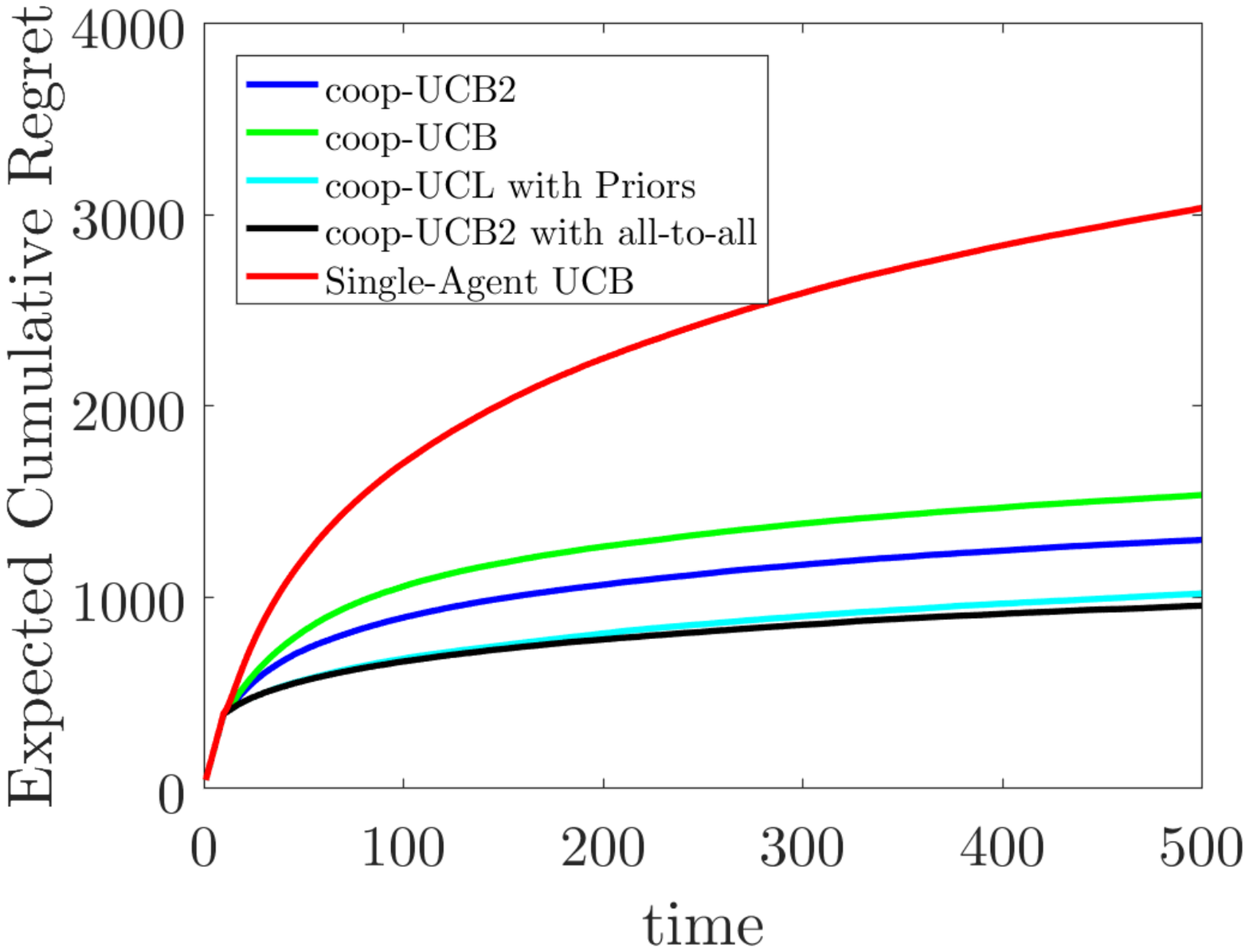}}\\

\subfigure[]{\label{ER_degcentsim}
	\includegraphics[width=.8\columnwidth]{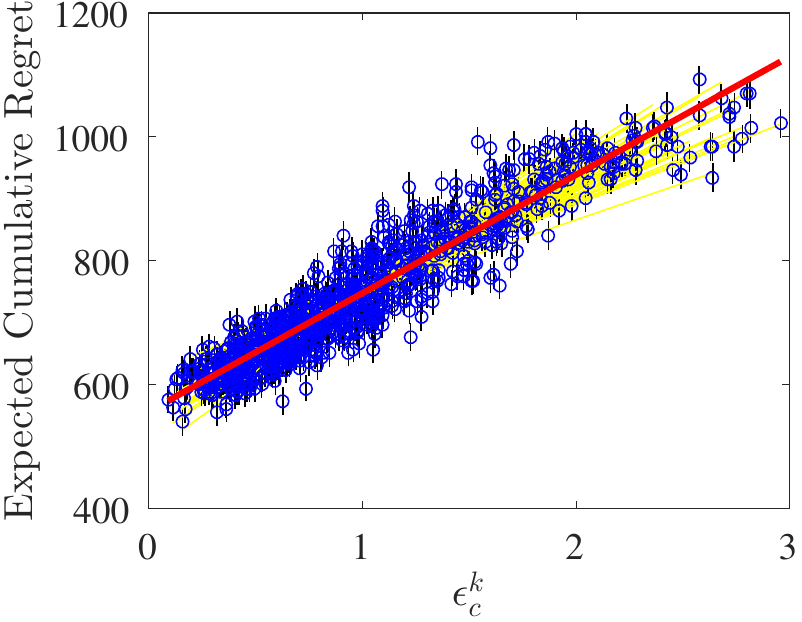}}

\caption{(a) Simulation results comparing expected cumulative regret for different agents in the fixed network shown, using $P$ as in \eqref{Pdefn} and $\kappa = \frac{d_{\text{max}}}{d_{\text{max}}-1}$. Note that agents $1$ and $2$ have nearly identical expected regret. (b) Simulation results of expected cumulative regret for several different MAB algorithms using the fixed network shown in  Fig.~\ref{5AgentGraphSim}. (c) Simulation results of expected cumulative regret as a function of normalized $\epsilon_c^k$ for nodes in ER graphs at $T=500$.  Also shown  in red is the best linear fit. }	\vspace{-0.1in}
\end{figure}

In this section, we elucidate  our theoretical analyses from the previous sections with  numerical examples. We first demonstrate that the ordering of the performance of nodes obtained through numerical simulations is identical to the ordering predicted by the nodal explore-exploit centrality measure: the larger the $\epsilon_c^k$ the lower the performance. We then investigate the effect of the graph connectivity on the performance of agents in random graphs. 

For all simulations we consider a $10$-armed bandit problem with mean rewards drawn from a normal random distribution with mean $75$ and standard deviation $25$. The sampling standard deviation  is $\sigma_s = 30$.  These parameters were selected to give illustrative results within the displayed time horizon, but the relevant conclusions hold across a wide variation of parameters. The simulations used $f(t)=\sqrt{\ln t}$.

%

\begin{example}[\bit{Regret on Fixed Graphs}]
Consider the set of agents communicating according to the graph in Fig.~\ref{5AgentGraphSim} and using the coop-UCB2 algorithm to handle the explore-exploit tradeoff in the distributed cooperative MAB problem. 
The values of $\epsilon_c^k$ for nodes $1,2,3,$ and $4$ are $2.31, 2.31, 0,$ and $5.43$, respectively. As noted in Remark~\ref{Remark:Indiv}, agent $3$ should have the lowest regret, agents $1$ and $2$ should have equal and intermediate regret, and agent $4$ should have the highest regret. These predictions are validated in our simulations shown in Fig.~\ref{5AgentGraphSim}. The expected cumulative regret in our simulations is computed using $5000$ Monte-Carlo runs.  

Fig.~\ref{MethodCompare} demonstrates the relative performance differences between coop-UCB, coop-UCB2, coop-UCL, and single agent UCB with the same run parameters.  Here the coop-UCL algorithm is shown with an informative prior and no correlation structure.  Each agent in the coop-UCL simulation shown here has $\Sigma_0 = 625 \cdot \mathcal{I}_M$ and $\boldsymbol{\nu}_0 = 75 \cdot \mathbf{1}_M$.  The use of priors markedly improves performance.
\end{example}




We now explore the effect of $\epsilon_c^k$ on the performance of an agent in an Erd{\"o}s-R{\'e}yni  (ER) random graph.  ER graphs are a widely used class of random graphs where any two agents are connected with a given probability $\rho$~\cite{bollobas1998random}.  

\begin{example}[\bit{Regret on Random Graphs}]
Consider a set of $10$ agents communicating according to an ER graph and using the coop-UCB2 algorithm to handle the explore-exploit tradeoff in the aforementioned MAB problem.  
In our simulations, we consider $100$ connected ER graphs, and for each ER graph we compute the expected cumulative regret of agents using $1000$ Monte-Carlo simulations  with $\rho = \ln(10)/10$,  $P$ as in \eqref{Pdefn}, and $\kappa = d_{\text{max}}/(d_{\text{max}}-1)$. We show the behavior of the expected cumulative regret of each agent as a function of the normalized $\epsilon_c^k$ in  Fig.~\ref{ER_degcentsim}.
It is evident that
increased $\epsilon_c^k$ results in a sharp decrease in performance.  Conversely, low $\epsilon_c^k$ is indicative of better performance.  This disparity is due to the relative scarcity of information at nodes that are in general less ``central.''  
\end{example}

%
%
%
%
%
%
%
%

\section{Final Remarks}
\label{FinalRemarks}
In this paper we used the distributed multi-agent MAB problem to explore cooperative decision-making in networks.
We designed the coop-UCB2 and coop-UCL algorithms, which are frequentist and Bayesian distributed algorithms, respectively, in which agents do not need to know the graph structure.   We proved bounds on performance, showing order-optimal performance for the group. 
Additionally, we investigated the performance of individual agents in the network as a function of the graph topology, using a proposed measure of nodal explore-exploit centrality. 

Future research directions include rigorously exploring other communications schemes, which may offer better performance or be more suitable for modeling certain networked systems.  It will be important to consider the tradeoff between communication frequency and performance as well as the presence of noisy communications.

\bibliographystyle{unsrt}
\bibliography{bibfile_abrev,bandits_abrev}

\end{document}